# Individually Addressable Nanoscale OLEDs

*Cheng Zhang\*, Björn Ewald\*, Leo Siebigs, Luca Steinbrecher, Maximilian Rödel, Monika Emmerling, Jens Pflaum\*, Bert Hecht\**

C. Zhang and B. Ewald contributed equally to this manuscript as joined first authors

C. Zhang, L. Siebigs, L. Steinbrecher, M. Emmerling, B. Hecht
Experimental Physics 5, University of Würzburg, Am Hubland, 97074 Würzburg, Germany
Email: cheng.zhang@uni-wuerzburg.de, bert.hecht@uni-wuerzburg.de

B. Ewald, M. Rödel, J. Pflaum
Experimental Physics 6, University of Würzburg, Am Hubland, 97074 Würzburg, Germany
Email: bjoern.ewald@uni-wuerzburg.de, jpflaum@physik.uni-wuerzburg.de

Keywords: organic light-emitting diodes, e-beam lithography, nanopixel, plasmonic nanoantennas, thermally activated delayed fluorescence, organic electronics, nanotechnology

**Abstract:** Augmented Reality (AR) and Virtual Reality (VR), require miniaturized displays with ultrahigh pixel densities. Here, we demonstrate an individually addressable subwavelength OLED pixel based on a nanoscale electrode capable of supporting plasmonic modes. Our approach is based on the notion that when scaling down pixel size, the 2D planar geometry of conventional organic light-emitting diodes (OLEDs) evolves into a significantly more complex 3D geometry governed by sharp nanoelectrode contours. These cause (i) spatially imbalanced charge carrier transport and recombination, resulting in a low quantum efficiency, and (ii) filament growth, leading to rapid device failure. Here, we circumvent such effects by selectively covering sharp electrode contours with an insulating layer, while utilizing a nano-aperture in flat areas of the electrode. We thereby ensure controlled charge carrier injection and recombination at the nanoscale and suppress filament growth. As a proof of principle, we first demonstrate stable and efficient hole injection from Au nanoelectrodes in hole-only devices with above 90 % pixel yield and longtime operation stability and then a complete vertical OLED pixel with an individually addressable nanoelectrode ($300 \times 300$ nm$^2$), highlighting the potential to further leverage plasmonic nanoantenna effects to enhance the performance and functionality of nano-OLEDs.

## 1. Introduction

In organic semiconductor technologies, vertical multilayer architectures offer precise control over optoelectronic properties, with applications ranging from organic light-emitting diodes (OLEDs)[1-3], to organic photodetectors[4] and vertical organic transistors[5]. The ongoing miniaturization of such components is a major driving force behind recent technological advancements. Examples include ultrasmall OLED pixels for Virtual Reality (VR) and Augmented Reality (AR) displays, as well as miniaturized transistor structures for lab-on-a-chip systems.[6-10] However, miniaturization often entails significant scaling effects, making it impractical to simply downscale devices without accounting for these phenomena.[11] With respect to organic devices, a few studies have addressed charge carrier transport and recombination in nanoscale junctions, but without utilizing individually addressable nano-electrodes.[12-14]

A key challenge in OLED miniaturization arises from sharp edges at nanoelectrodes. The resulting locally enhanced electric fields cause modified Schottky barriers at metal-semiconductor contacts and correspondingly modified charge-carrier injection/extraction mechanisms. For example, the dominant tunneling contribution to the overall charge carrier injection for diodes < 100 nm is known to impose severe limitations on the device performance for materials with low charge-carrier mobilities such as organic semiconductors.[15-17] Local electric field enhancement is expected to cause charge injection barrier minima, which will lead to current density hotspots and charge imbalance throughout the device (see **Figure 1**a).[18] In addition, local electric field enhancement typically leads to metallic filament formation, associated with instable and non-deterministic device operation (see Figure 1a).[19-21] Also, the power emitted by an OLED is unfavorably affected by miniaturization since for small pixels it scales with $(l/\lambda)^2$, where $l$ is the characteristic pixel dimension and $\lambda$ is the free-space wavelength, whereby the emitted power quickly drops for subwavelength devices.[22] An intuitive solution makes use of resonant plasmonic antennas as passive scatterers integrated into standard OLEDs to enhance radiation.[23-26] We recently reported the integration of active subwavelength plasmonic nanoantenna electrodes into laterally arranged nano-OLEDs and demonstrated enhanced emission by means of antenna effects.[27] However, devices suffered from local-field-induced filament growth and subsequent premature device failure, and state-of-the-art multilayer organic stack designs with all their benefits could not be employed.[27, 28]

Here, we introduce a new approach combining advantages of vertical stacking with potential benefits of nanoscale plasmonic metal electrodes yet avoiding the adverse effects of sharp edge structures associated with miniaturization. The idea is summarized in Figure 1.

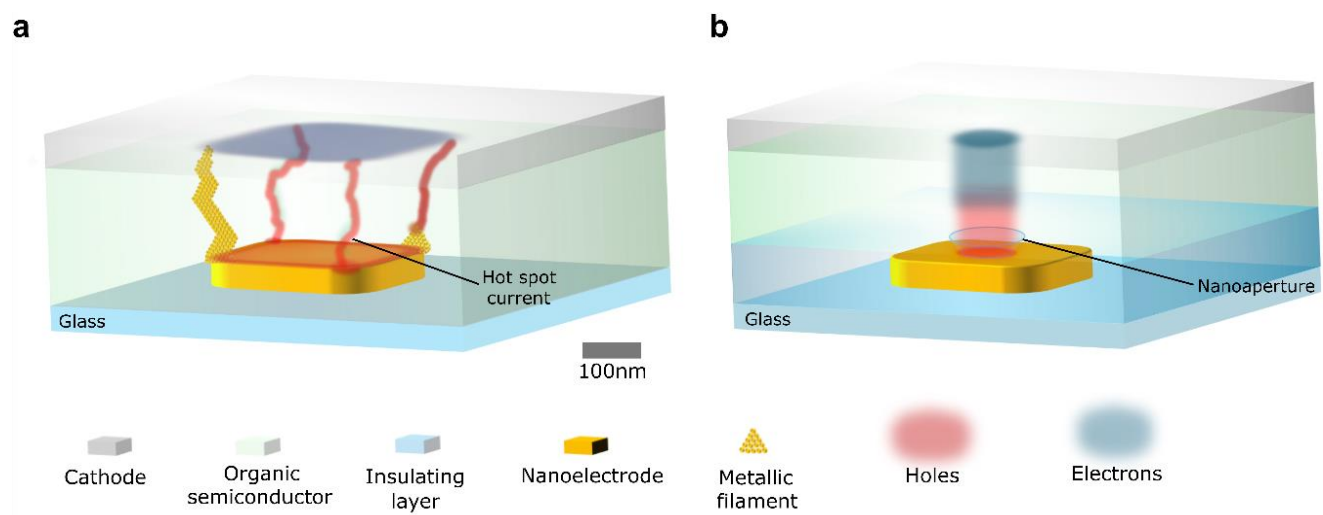

**Figure 1.** Schematic presentation of current pathways in individually addressable vertically stacked metal-organic-metal optoelectronic devices with nanoscale bottom electrodes. The organic stack is sandwiched between a hole-injecting Au nanoelectrode, and an electron-injecting extended top electrode (cathode). The electron (hole) current density distribution is highlighted in blue (red). a) Device configuration without edge and corner insulation where electric field enhancement generates hotspots for hole-injection and metallic filament growth. This leads to a drastically reduced recombination efficiency and to device instability. b) Device configuration with edge and corner insulation. The hole-injection from the nanoelectrode is limited to a nanoaperture centered at an area of homogenous electric field thus bypassing the detrimental effects of electrode downscaling.

We employ a nanoaperture to constrain hole-injection from a nanoelectrode to its planar center by covering its peripheral corners and edges with an insulating layer (Figure 1b). This effectively mitigates the detrimental effects caused by electric field hot spots at these regions (Figure 1a), which results in stable device operation with balanced charge carrier transport and recombination dynamics. We validate this approach by first demonstrating nanoaperture-controlled hole injection through single Au nanoelectrodes in hole-only devices, followed by the demonstration of unprecedentedly small, individually addressable, vertically-stacked OLED nanopixels with quantum efficiencies in the 1 % range indicating balanced charge transport and recombination. Our methodology represents a significant advance in the miniaturization of optoelectronic devices by demonstrating a precise spatial control of the charge carrier paths and thus, of the active recombination and emissive regions in nanoscale OLED architectures as well as by paving the road towards leveraging additional plasmonic nanoantenna effects in Au nanoelectrodes. As such, it holds substantial implications, e.g., for the development of next-generation ultrasmall but high-resolution displays and other nano-optoelectronic systems.[29-31]

## 2. Results and Discussion

We first discuss the implementation of hole-only devices based on $1 \times 1$ µm$^2$ Au patches. To implement complete OLEDs, we further downscale individually addressable electrode patches to $300 \times 300$ nm$^2$, to allow for improved light outcoupling via decay of the nanoelectrode's plasmonic modes into photons.

### 2.1. Au Nanoelectrodes with Nanoaperture

Electrostatic simulations of a $1 \times 1$ µm$^2$ quadratic electrode patch (**Figure S1**) suggest a 3-fold electric field enhancement at the electrode edges and a 6-fold electric field enhancement at the corners compared to the electrode center. $1 \times 1$ µm$^2$ quadratic Au nanoelectrodes are fabricated using standard electron beam lithography (EBL) and thermal metal evaporation. To achieve a uniform electric field away from edges and corners that enables controlled charge injection, a high-quality thin metal film with minimal surface roughness is crucial. We employ an evaporation rate of $1.5$ nm·s$^{-1}$ in a high vacuum environment to deposit a 50 nm Au film resulting in a root mean square (rms) surface roughness of 1 nm (see **Figure S2**). To insulate edges and corners, a second EBL process utilizing the high-resolution negative resist hydrogen silsesquioxane (HSQ)[32], is used to create an insulating layer with a nanoaperture on top of the Au nanoelectrode, as shown in **Figure 2**.

To this end, a gradient electron beam dose is applied across the entire antenna pixel, ranging from zero at the electrode center to full dosage at the electrode edges (Figure 2a). This results in a precise control of the shape and depth of the resulting central nanoaperture after the subsequent removal of unexposed HSQ resist, using tetramethylammonium hydroxide (TMAH). A successful opening of a nanoaperture is confirmed by tapping mode atomic force microscopy (AFM), as displayed in Figure 2b and 2d. Additionally, the remaining electrical connection pad is insulated to prevent any potential leakage currents. To achieve a clean, open nanoaperture, a mild oxidative etching step is applied to the exposed Au portion of the electrode, using a highly diluted Lugol's solution. Furthermore, conductive AFM measurements are conducted, as illustrated in Figure 2c, which show that the conductive area coincides precisely with the nanoaperture.

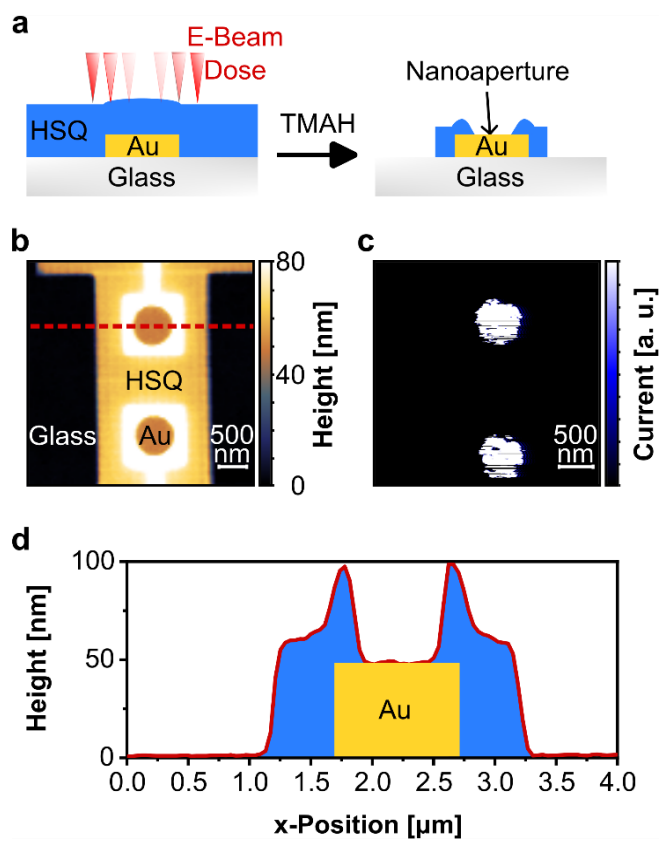

**Figure 2.** Au nanoelectrodes with nanoapertures: a) EBL process with HSQ (blue) as negative resist covering the electrode. Crosslinking is controlled by a gradient E-beam dose indicated by the fading red color. Alkalic development with tetramethylammonium hydroxide (TMAH) results in the removal of unexposed HSQ and partial etching of exposed HSQ according to the degree of crosslinking. b) Tapping-mode AFM images of two adjacent $1 \times 1$ µm$^2$ Au electrodes with electrical connectors. The electrode edges and the connector are fully covered by HSQ apart from a central 550 nm diameter nanoaperture. c) Conductive AFM image of the same area as in b revealing current injection and flow only inside the apertures. d) Cross sectional height profile along the cut indicated by the dashed red line in b).

## 2.2. Hole-Only Devices

To confirm the operation of the fabricated Au electrode patches with nanoaperture opening as hole-injecting electrodes we have studied hole-only devices with a hole-injecting anode and an electron-blocking cathode, which are expected to provide unipolar hole transport through the organic semiconductor material. The performance of nanojunctions fabricated based on the electrode patches of Figure 2 is compared to conventional macrojunctions, with an active area of $100 \times 100 \, \mu m^2$. The active area of the nanojunctions ($2.4 \cdot 10^{-9} \, cm^2$) is five orders of magnitudes smaller than that of the macrojunction ($1.0 \cdot 10^{-4} \, cm^2$), and edge effects are completely negligible in the macrojunction. The general structure of the hole-only devices and a juxtaposition of macro- and nanojunction properties, are depicted in **Figure 3**.

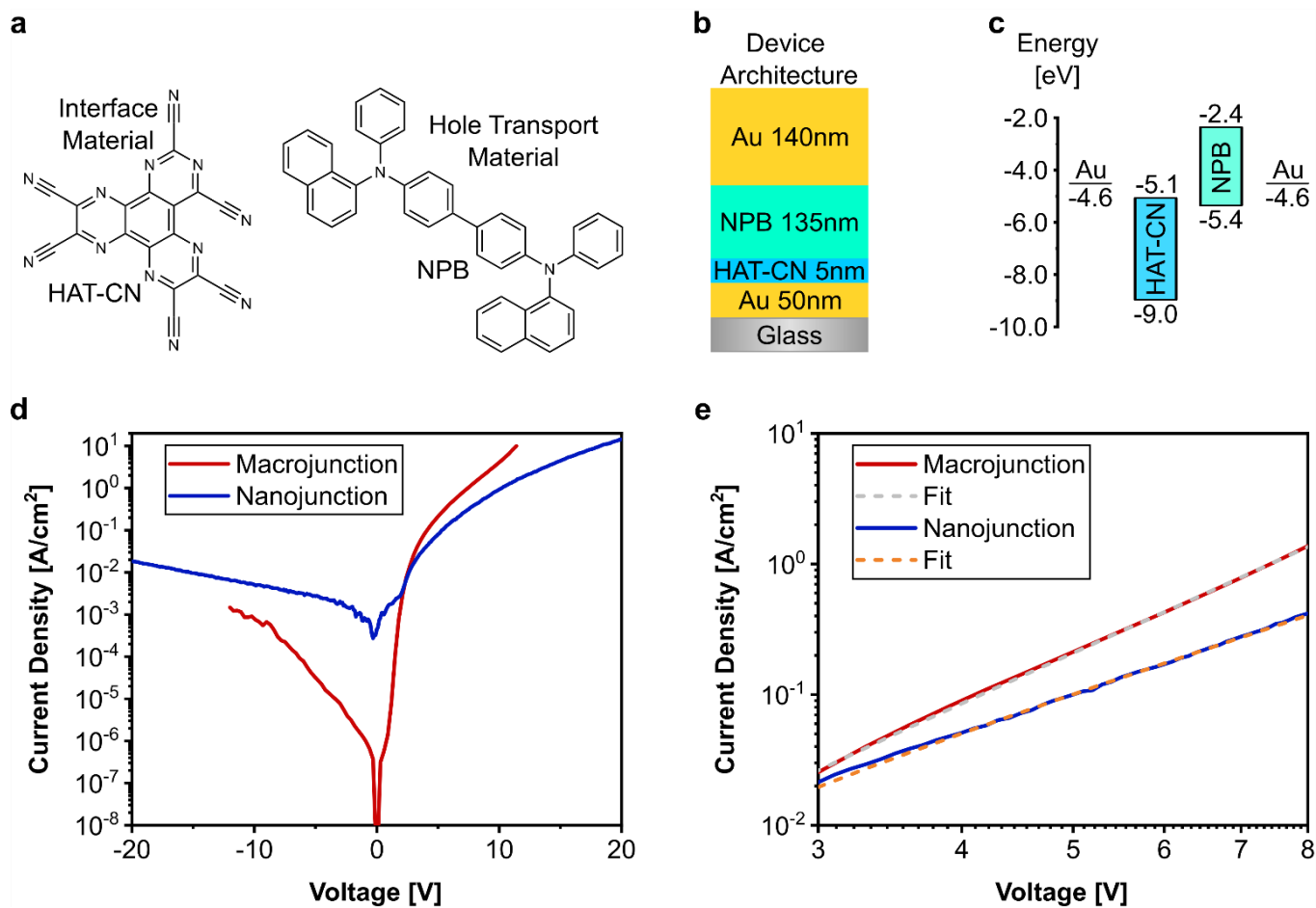

**Figure 3.** Electrical properties and device configuration of hole-only macrojunctions (electrode patch: $100 \times 100 \, \mu m^2$, active area: $1.0 \cdot 10^{-4} \, cm^2$) and nanojunctions (electrode patch: $1 \times 1 \, \mu m^2$, nanoaperture diameter: 550 nm, active area: $2.4 \cdot 10^{-9} \, cm^2$): a) Molecular structures of the interface functionalization material HAT-CN and the hole transport material NPB. b) Device architecture consisting of an Au bottom anode (50 nm thickness) functionalized with an ultrathin HAT-CN layer (5 nm), followed by a hole-transport layer consisting of NPB (135 nm) and the electron-blocking Au cathode (140 nm). c) Resulting flat-band energy landscape without considering Fermi-level pinning of HAT-CN and NPB. d) Semilogarithmic current density-voltage characteristics of representative macro- and nanojunction pixels. e) Corresponding double-logarithmic presentation of the space-charge limited current and the Poole-Frenkel fit in the regime between 3 and 8 V.

N,N′-Di(1-naphthyl)-N,N′-diphenyl-(1,1′-biphenyl)-4,4′-diamine (NPB, molecular structure see Figure 3a) has been chosen as organic hole transport material for its well-established electronic properties and its use in many standard OLED applications.[33] The device stack architecture and the flat-band energy landscape are displayed in Figure 3b and 3c. The work function of the polycrystalline Au electrodes is expected to be in the range of -4.4 to -4.7 eV.[34] The highest occupied molecular orbital (HOMO) level of NPB is at -5.4 eV and the lowest unoccupied molecular orbital (LUMO) level is at -2.4 eV.[35] The mismatch of the LUMO level of NPB with the work function of Au makes it feasible to use Au also as an electron blocking top electrode. 1,4,5,8,9,11-Hexaazatriphenylenehexacarbonitrile (HAT-CN, molecular structure see Figure 3a) is used to functionalize the Au bottom contact. The energy level alignment across the Au/HAT-CN/NPB interface is determined by the Fermi-level pinning of the HAT-CN LUMO and the NPB HOMO.[36] The overall device stack architecture (Figure 3b) comprises Au(50 nm)/HAT-CN(5 nm)/NPB(135 nm)/Au(140 nm).

The effectiveness of the HAT-CN interface functionalization is demonstrated by the current density-voltage (*J-V*) characteristics recorded for a macrojunction device with and without HAT-CN functionalization (**Figure S3**). Upon HAT-CN functionalization an up to six orders of magnitude increase in hole current at 10 V with a small onset voltage of only 0.9 V for hole injection is observed. Kahn et al.[37] have described the formation of an interfacial dipole of 1.2 eV at the Au/NPB interface, leading to an additional increase of the hole-injection barrier. The functionalization by an ultrathin HAT-CN layer hence results in a drastic reduction of the hole-injection barrier due to Fermi-level pinning and the related interface dipole, making our polycrystalline Au electrodes a superb platform for hole-injection in the following device applications.

Figure 3d shows a comparison of the *J-V* characteristics between the nano- and macrojunction. The deviation of the current densities at 0 V between macro- and nanojunction is related to operation of the nanojunction below the instrumental resolution of the source measurement unit of around 1 pA. The estimated onset voltage for hole-injection is 0.9 V for the macrojunction and 1.9 V for the nanojunction, respectively. The slight difference can be explained by the larger probability for hole-injection and current hotspots in the macroscopic device, as well as by different processing conditions of the Au electrodes, which may alter the work function of the polycrystalline Au surface and the hole-injection barrier.[18, 38] The nanojunction is stable in a voltage regime between 20 V to -20 V, while the macrojunction is only stable between 12 V to -12 V exhibiting electrical breakdown for higher voltages. We associate the higher stability of the nanojunction as compared to the macrojunction with the fact that the probability for electrode defects and filament formation is proportional to the device area. It also further signifies the efficiency of the edge and corner blocking process described above.

At 10 V, current densities of 1 A·cm$^{-2}$ and of 4 A·cm$^{-2}$ are reached for the nanojunction and for the macrojunction, respectively. The rather small deviation corroborates the cleanliness of the active area after nanoaperture fabrication and oxidative cleaning. The higher current density in the macrojunction likely occurs due to inhomogeneous charge transport via low-ohmic pathways.[39] As directly visible from **Figure S4** the nanojunction operates with an absolute current of only 10 nA at 10 V, which is 5 orders of magnitude below that of the macrojunction (1 mA).

Efficient hole-injection via the Au nanoelectrode is a prerequisite for efficient light emission in OLED structures, where current densities in the range of 1·10$^{-3}$ to 1·10$^{-1}$ A·cm$^{-2}$ at low operating voltages are desirable. The blocking ratio, defined by the current density ratio in forward and reverse bias direction, is 8·10$^{2}$ (@ 20 V) for the nanojunction in relation to 5·10$^{3}$ (@ 10 V) for the macrojunction. We attribute the lower blocking ratio for the nanojunction to a slight curvature of the top Au contact induced by the surface topology of the nanoaperture underneath (Figure 2b and 2d). This is supported by *J-V* characteristics of hole-only nanojunctions with varying hole diameters (**Figure S5**). A smaller nanoaperture diameter is supposed to impose a more pronounced curvature at the top contact, resulting in a higher leakage current translating into a lower blocking ratio of the respective *J-V* characteristics. We have fitted the *J-V* characteristics based on the assumption of a space-charge limited current model in combination with a Poole-Frenkel-type charge carrier transport, the latter considering e.g. trapping and electrical-field assisted release of charge carriers within the organic semiconductor (**Equation 1**). In Equation 1, $\varepsilon_0$ is the vacuum permittivity, $\varepsilon_r$ is the dielectric constant of NPB, $\mu_0$ is the zero field mobility, *V* is the voltage corrected for the built-in voltage, *d* is the organic-stack thickness and $\beta$ is the Poole-Frenkel parameter.[1, 40]

$$J = \frac{9}{8}\varepsilon_0\varepsilon_r\mu_0 \cdot \frac{V^2}{d^3} \cdot exp\left(0.89 \cdot \beta \cdot \sqrt{\frac{V}{d}}\right) \qquad (1)$$

With $\mu_0$ and $\beta$ as free parameters, the resulting fit functions agree well with the measured *J-V* characteristics in a range between 3 to 8 V, for both the macrojunction and the nanojunction (Figure 3e). We restrict the voltage regime to 3 to 8 V, to avoid influences by tunnel injection, deep-trap filling or Joule heating. The zero-field hole mobility amounts to $1 \cdot 10^{-5}$ $cm^2 \cdot V^{-1} \cdot s^{-1}$ for the macrojunction, and $3 \cdot 10^{-5}$ $cm^2 \cdot V^{-1} \cdot s^{-1}$ for the nanojunction. The mobility values are in good agreement with literature values of NPB in the presence of trap states.[41] The Poole-Frenkel parameter $\beta$ converges to $5 \cdot 10^{-3}$ for the macrojunction and to $2 \cdot 10^{-3}$ for the nanojunction, respectively. Both, the lower hole mobility and the smaller Poole-Frenkel parameter $\beta$ can be explained by a lower absolute number of trap states and, thus, lower influence on the hole carrier transport within the ultrasmall active volume of the nanojunction.

To substantially illustrate the benefits of the nanoaperture concept for the stability and efficiency of devices, we have performed a comparative study of the electrical properties for hole-only devices based on Au nanoelectrodes ($1 \times 1$ $\mu m^2$) with and without nanoaperture (see **Figure 4**).

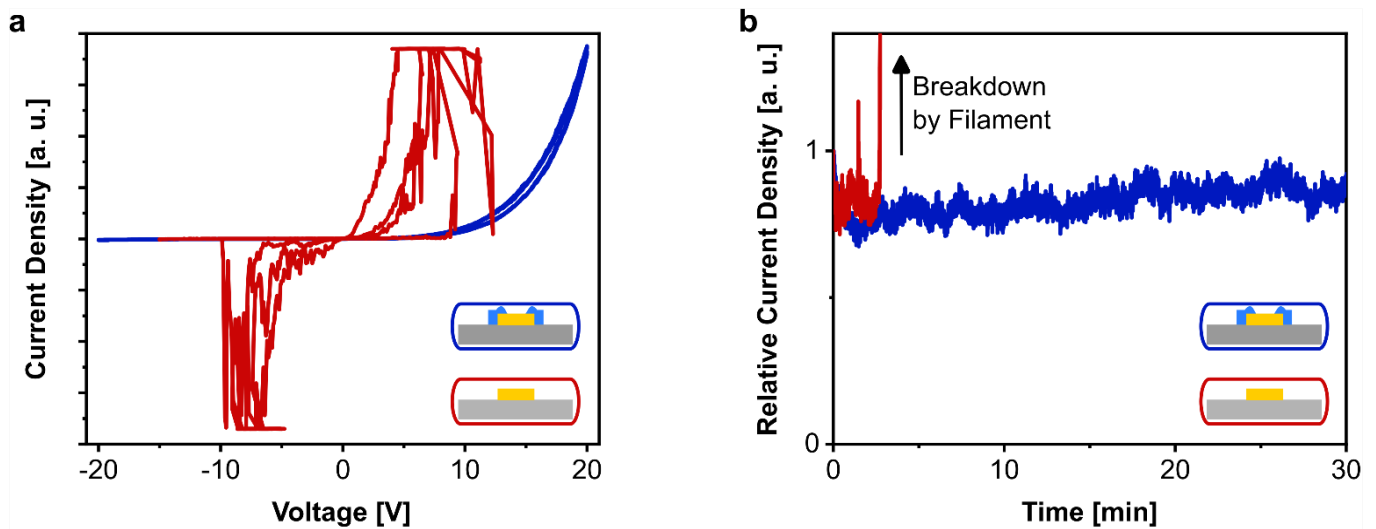

**Figure 4.** Electrical properties of hole-only devices based on Au nanoelectrodes ($1 \times 1$ $\mu m^2$) with nanoaperture opening (blue lines and inset) and without edge insulation (red lines and inset): a) Current density-voltage characteristics of two representative junctions with and without nanoaperture. The junctions have been voltage-cycled three times starting from 0 V. An erratic behavior, likely resulting from filament formations and disruptions are observed for a nanoelectrode without edge insulation. b) Constant voltage operation (5 V, dc) of two representative junctions with and without nanoaperture. The device without nanoaperture

exhibits breakdown by filament formation after only three minutes of operation, while the device with nanoaperture is stable over the measurement period of 30 minutes.

The *J-V* cycles (Figure 4a) demonstrate the effectiveness of the edge insulation in an impressive manner. While the electrodes with nanoaperture display deterministic device operation, as discussed in detail above, the electrodes without edge insulation show erratic behavior during cycling. The abrupt jumps to the upper current limitation are attributed to Au atom migration with subsequent low-ohmic filament formation, which is incited by the up to 3-fold electric field enhancement at electrode edges and up to 6-fold electric field enhancement at the electrode corners (Figure S1). Upon applied reversed voltage, the Au filaments are typically disrupted, and the overall current is again dominated by low-ohmic pathways. Our device with nanoaperture electrode does not show any tendency for filament formation and exhibits only a small hysteresis in the *J-V* cycles visible in forward bias direction. Taking into account that the reported *J-V* cycles are scaled for comparability, we note that as expected the absolute current through the nanoaperture junction is smaller than for the junction without edge insulation.

To further test the device operation stability, we have applied a constant dc voltage of 5 V. The time dependence of the relative current density is depicted in Figure 4b. The electrode without edge insulation shows electrical breakdown after only three minutes of operation. Remarkably the relative current density of the nanoaperture device is fully stable, at least, over the measurement period of 30 minutes.

We further observe that devices with nanoapertures exhibit an outstanding reproducibility as demonstrated by the *I-V* characteristics in **Figure 5**, where the blue area marks the variation of the current for 30 individual junctions. The *I-V* curve highlighted in red, corresponds to the nanojunction discussed in Figure 3. The current variations remain within one order of magnitude for forward and reverse bias @ 20 V, indicating only small variations in size and depth of the nanoaperture as well as a high quality of the Au interface after processing. Correspondingly, 91 % (30 out of 33) of the fabricated nanojunctions have shown no tendency for filament formation, again clearly demonstrating the advantages of effective edge coverage and surface quality in nanoaperture devices.

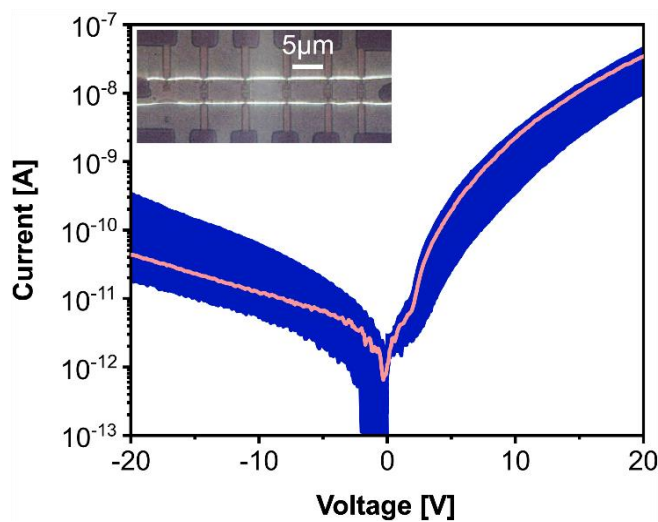

**Figure 5.** Absolut variation of current voltage characteristics over 30 individual nanojunctions illustrated by the blue area. The current variation between the investigated 30 nanojunctions remains within one order of magnitude @ 20V. The *I-V* curve highlighted in red corresponds to the nanojunction in Figure 3. The inset shows a white-light transmission micrograph of the electrode layout with 11 pixels in one block. Three blocks (33 pixels) of the same configuration were fabricated on one sample, with only three pixels featuring an anomalous behavior.

Since the final goal of using Au electrodes is to miniaturize towards resonant plasmonic nanoantennas we also investigated smaller electrode patches and corresponding apertures. We found that the concept is equally effective for hole-only nanojunctions with an electrode patch size of only $300 \times 300 \ nm^2$ and a nanoaperture diameter of 200 nm (**Figure S6**). The *J-V* characteristics (Figure S6a) are comparable to that of the $1 \times 1 \ \mu m^2$ electrode patches. The observed lowering of the blocking ratio is associated with a smaller nanoaperture diameter, without having a negative impact on the device stability under dc voltage operation (Figure S6b). We therefore conclude that our approach is excellently suited for the prospective integration of resonant plasmonic nanoelectrodes with subwavelength dimensions into vertical organic electronic devices.

## 2.3. Nanoscale OLED Pixels

To apply the nanoaperture concept to an actual light-emitting device, we have fabricated OLED nanopixels. These nanopixels constitute, to best of our knowledge, the first individually addressable subwavelength nano-OLED pixels in a standard vertical multilayer architecture. To enhance light outcoupling into the 1 % range, we have further reduced the pixel size to $300 \times 300$ nm$^2$. Yet, so far, the opaque character and the multimode plasmonic resonances of the Au anode limit the actual external quantum efficiency of the vertical OLED nanopixel as compared to structures exhibiting a single radiative plasmonic mode.[27]

An overview of the OLED architecture and the resulting device properties is provided in **Figure 6**. The flat-band energy diagram representing the relevant transport and recombination levels in the device is displayed in Figure 6a. The bottom anode consists of a 50 nm thick Au nanoelectrode ($300 \times 300$ nm$^2$) with a nanoaperture opening (200 nm in diameter). Hole injection is mediated by a 5 nm HAT-CN interface layer followed by 30 nm of NPB acting as hole-transport layer as described above. As light-emitting material, we make use of the thermally-activated delayed fluorescence (TADF) emitter 2-[4-(diphenylamino)phenyl]-10,10-dioxide-9H-thioxanthen-9-one (TXO-TPA) embedded in 1,3-bis(N-carbazolyl)benzene (mCP) as host material (7 vol%), which constitutes an efficient and commonly-used host-emitter combination.[42] The emissive layer (30 nm) is separated by a 75 nm bathophenanthroline (Bphen) electron-transport layer from the top electrode. The flat top cathode consisting of 10 nm of Ca and 120 nm Al supports electron injection. The asymmetric device architecture is intended to achieve exciton recombination far away from the top cathode to suppress extra light absorption by the cathode layers.

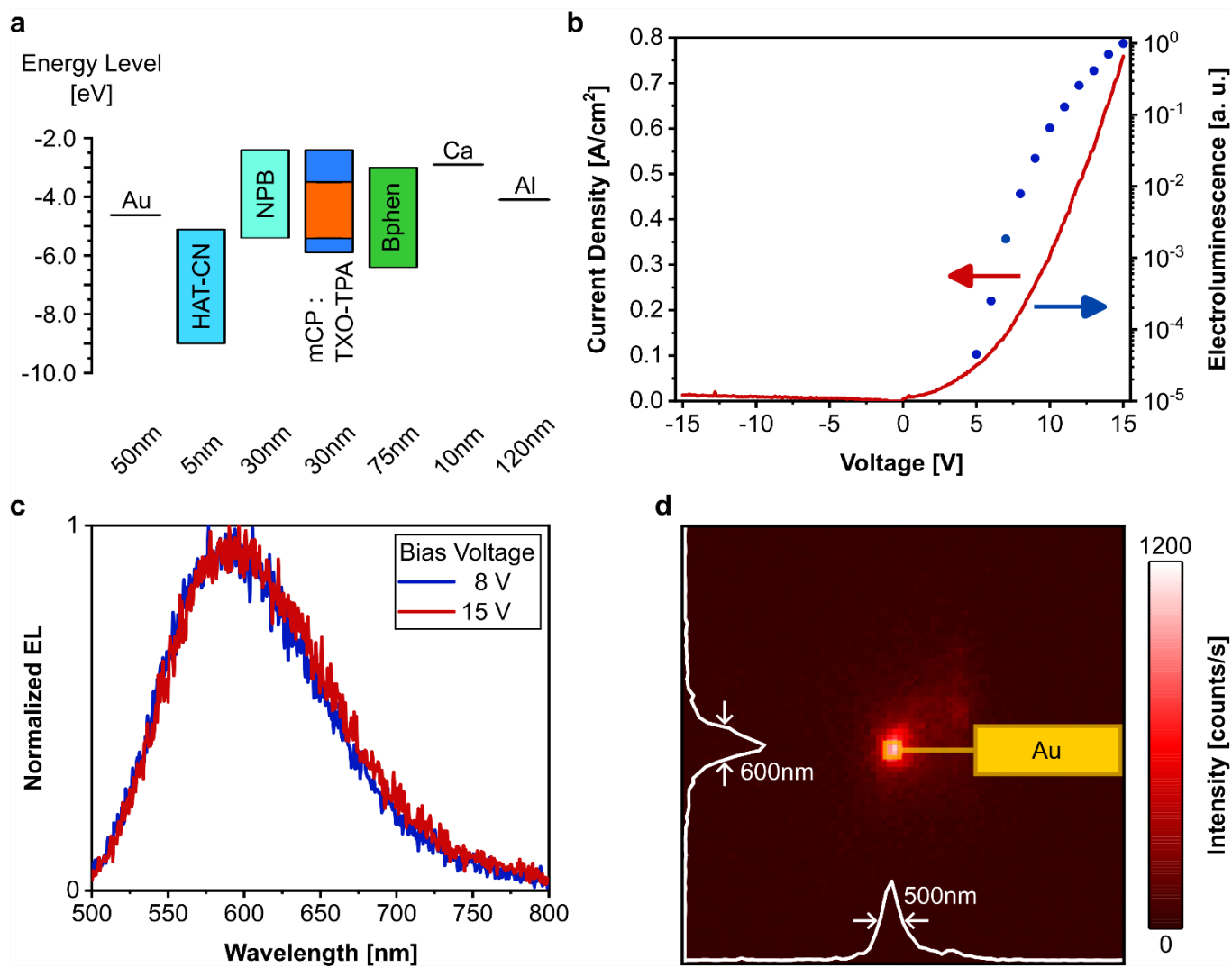

**Figure 6.** Nano-OLED device based on Au nanoelectrodes ($300 \times 300$ nm$^2$) with a nanoaperture of 200 nm in diameter: a) Flat band energy landscape and device architecture, with HAT-CN (5 nm) as injection layer, NPB (30 nm) as hole transport layer, mCP doped with TXO-TPA as emissive layer (30 nm), Bphen (75 nm) as electron transport and hole blocking layer, Ca (10 nm) as injection layer and Al (120 nm) as capping electrode. b) Current density-voltage-electroluminescence characteristics (-15 V to 15 V). c) Normalized EL spectra measured at 8 V and 15 V constant voltage operation. d) Map of the spatial distribution of emitted light from a representative pixel. The emission is centered at the pixel area and defined by the nanoaperture. The electrode structure and its electrical connection wire are superimposed in yellow color. The intensity linecuts to the left and at the bottom reveal a point-like emission pattern with a full-width-at-half-maximum below 600 nm.

Remarkably, the fabricated OLED nanopixels exhibit a stable device operation under forward and reverse bias with a pronounced *J-V* asymmetry (Figure 6b). The pixels can be cycled under forward and reverse bias without device failure (see **Figure S7a**). The onset of light emission occurs at 5 V and the recorded electroluminescence intensity then increases with increasing voltage (Figure 6b). In accordance with our previous observations on nanoaperture-based hole-only devices, no tendency for filament formation, subsequent short circuits and, hence, device failure is observed, again highlighting the inalienability of the nanoaperture concept when building devices based on individually addressable nanoscale electrode patches. The electroluminescence spectrum of the OLED nanopixel is reported in Figure 6c for constant-voltage operation at 8 V and 15 V, respectively. The EL spectrum is centered around 580 nm, which coincides with the molecular TADF emission of TXO-TPA in mCP.[42] The spectral shape and position are independent of the voltage applied which indicates the spatial stability of the narrow emission zone and, thus, the absence of cavity effects in the pixel emission.

The far-field microscopy image of the spatial distribution of the electroluminescence is displayed in Figure 6d with the actual pixel dimension and the electrical connections superimposed and marked in yellow color. As inferred from the intensity linecuts a full-width-at-half-maximum of the emission spot of less than 600 nm is observed. Despite the non-transparent character of the Au electrode and the concomitant presence of multiple higher-order plasmonic modes, we achieve remarkably high external quantum efficiencies of up to 1 % (Figure S7b). Such a high value indicates a high internal recombination efficiency, which is expected because of the used high-performance material stack in combination with the nanoaperture, which enforces balanced charge carrier transport also in nanoscale devices.

## 3. Conclusion

We have identified the inhomogeneous 3D electric field distribution at nanoelectrodes as the primary reason for spatially unbalanced charge carrier injection and transport as well as device failure due to filament formation in vertical nano-optoelectronic devices. To circumvent these detrimental scaling effects, we propose a novel concept utilizing strategically designed insulating layers that specifically cover the edges and corners of metallic nanoelectrodes while leaving a nanoscale aperture in flat parts of the nanoelectrode. The implementation of such a nanoaperture has proven to be essential for the long-term stability of devices. Accordingly, by effectively restricting the charge injection area to a plane interface of homogenous electric field distribution, erratic *J-V* characteristics, directly related to electric field induced filament formation at metal electrode edges of hole-only devices with bare Au nanoelectrodes ($1 \times 1$ µm$^2$) can be completely suppressed. Our nanofabrication procedure is highly reproducible with a pixel yield above 90 % and thereby paves the way for the realization of organic (opto-)electronic devices with individual nanoscale electrodes. We have demonstrated the working principle of a vertical metal-organic nanodevice with an individual controllable pixel. In hole-only devices, Au nanoelectrodes with nanoaperture and HAT-CN functionalization exhibit turn-on voltages as low as 1.9 V for hole injection and a remarkable device stability up to 20 V outcompeting their macroscopic counterpart with an electrode size of $100 \times 100$ µm$^2$. As a further proof-of-concept we have integrated a nanoaperture electrode into a standard vertical OLED architecture with individual addressable pixels of $300 \times 300$ nm$^2$, which is to best of our knowledge the smallest pixel size so far reported for OLEDs. The OLED nanopixels exhibit a low turn-on voltage of 5 V and stable operation between -15 to 15 V. Despite the non-transparent character of our nanoelectrodes, which represent plasmonic scatterers supporting mostly weakly-radiative higher-order modes, we achieve external quantum efficiencies of up to 1 % indicative of a large internal quantum efficiency. More advanced organic stack concepts and/or resonant plasmonic outcoupling elements supporting a single radiative mode, are expected to lead to further reduced pixel sizes and to an even more localized spatial emission in addition to an even higher external quantum efficiency.

## 4. Experimental Section

*Fabrication of Hole-Only Macrojunctions*: Glass substrates (Karl Hecht, $22 \times 22$ mm$^2$, thickness 170 µm) were thoroughly cleaned by consecutive ultrasonication (15 minutes each) in double distilled water (CarlRoth) with mucasol® detergent (schuelke), in double distilled water, in acetone and in isopropanol, and were dried in a nitrogen stream. The active area of the junction was defined by the overlap area of a perpendicular bottom and top contact stripe. The width of the stripes was defined to 100 µm each by stainless steel shadow masks (Beta LAYOUT GmbH) resulting in a junction area of 100 µm$^2$. The metal contacts and the organic films were deposited in two separate vacuum chambers, to avoid any contamination of the organic materials. In both chambers the deposition rate and thickness were controlled with a quartz crystal microbalance. Au was purchased at a purity of 99.99 % and evaporated from commercial molybdenum boat sources (Kurt J. Lesker) at a base pressure below $3 \cdot 10^{-6}$ mbar and at a deposition rate of 1.5 nm·s$^{-1}$. The Au bottom contact was deposited in a thickness of 50 nm. During transfer to the organic chamber the bottom contact was exposed to ambient conditions under yellow light for a short period of 5 to 10 minutes. The organic materials were evaporated from boron nitride crucibles by resistive heating at a base pressure of $10^{-9}$ mbar. HAT-CN and NPB were purchased from Ossila at double sublimed and sublimed grade, respectively, and were used without further purification. Both materials were deposited through a large area shadow mask widely covering the bottom contact. 5 nm HAT-CN was deposited at a deposition rate of 2 nm·min$^{-1}$ and 135 nm NPB was deposited at a deposition rate of 8 nm·min$^{-1}$. 140 nm NPB were deposited for the reference device without HAT-CN. After an ambient exposure of 10 minutes for sample transfer, the Au top contact was deposited within the metal chamber at a thickness of 140 nm. Finally, the devices were encapsulated in a glovebox (Jacomex, 0.0 ppm $O_2$, 0.0 ppm $H_2O$) with a glass slide and epoxy resin (Loctite EA9492 LI), to exclude the penetration of oxygen and water during the measurements.

*Fabrication of Hole-Only Nanojunctions and OLED Nanopixels*: The electrode layout was fabricated using optical lithography (AR-U4030, Allresist GmbH). Once developed, the samples were washed with water and plasma cleaned (200 W, 60 s, 10 sccm oxygen). Next, a 10 nm chromium adhesion layer followed by a 70 nm Au layer were deposited using electron beam physical vapor deposition at a rate of 1 nm·s$^{-1}$. The photomask was removed by a lift-off process in acetone. The electrode layout substrate was further cleaned in a series of steps using deionized water, acetone, and IPA for 10 minutes each in an ultrasonic bath. In order, to guarantee a completely residue-free substrate layout, 10-minute oxygen plasma cleaning was carried out (power 250 W, oxygen flow 20 sccm). To achieve high-resolution electron beam patterning on the layout substrate, a double-layer consisting of a 100 nm PMMA 600 K and a 20 nm PMMA 950 K (Allresist GmbH) was applied by spin-coating. The PMMA layers were then tempered on a hotplate at 150 °C for 3 minutes. Subsequently, a conductive resist (Electra 92, Allresist GmbH) was spin-coated onto the PMMA double layer to serve as a discharging layer for electron-beam lithography. The electron-beam writing process is then carried out using a Zeiss SEM Gemini 450 with an acceleration voltage of 30 kV and a beam current of 35 pA. The typical dose for the antenna pattern is around 800 µC·cm$^{-2}$. The development of the electron-beam resist was executed at room temperature following these steps: 1) Immersion of the sample in water for 20 s to remove the conductive resist, 2) Soaking in a mixture of MIBK and IPA (3:1 ratio) for 60 s to define the PMMA pattern, 3) 30 s immersion in IPA to terminate the development process, 4) Drying with a stream of nitrogen gas. A 50 nm Au thin film was then thermally evaporated at a deposition rate of 1.5 nm·s$^{-1}$ at a base pressure below 3·10$^{-6}$ mbar. Afterwards the sample was immersed into acetone (p. a. grade) overnight and 10 s of ultrasonic agitation were applied to complete the lift-off procedure of the PMMA mask. A layer of 100 nm HSQ (SX AR-N 8200, Allresist GmbH) was spin-coated along with a conductive resist layer to create a planarization layer. The HSQ layer was then structured using a second electron-beam writing step, with a gradient electron-beam dose used for the antenna patch (ranging from 0% to 100% of 2250 $\mu C \cdot cm^{-2}$) and a constant dose of 2250 $\mu C \cdot cm^{-2}$ for the remaining insulating layer. Development was carried out using a solution of 1 % tetramethylammonium hydroxide (TMAH) in water at room temperature. An AFM scan was performed on the structured antenna to ensure the quality of the development. To prevent any direct or weak electrical connections between the bottom Au connector and the top metal layer, it is necessary to add a thick insulating layer. To this end, a thick PMMA layer (~800 nm) was spin-coated on the sample and a third electron-beam writing was applied to define an open window channel aligned with the antenna electrode area for the pixel definition. After completion of the last electron-beam lithography step, an oxidative cleaning step with highly diluted Lugol´s solution (1:2000) was applied for 8 to 10 s with a consecutive washing step in water. For organic deposition a stainless-steel shadow mask (Beta LAYOUT) with an opening of 140 x 100 $\mu m^2$ was aligned on top of the nanoelectrode array. For metal top contact deposition, a shadow mask with a slightly larger opening of 200 x 120 $\mu m^2$ was used and aligned after completion of organic deposition. The shadow masks were fixed with tension strings and an ambient exposure of 45 minutes was needed for mask alignment. Apart from that, the organic deposition, the Au top contact deposition, and the encapsulation were carried out with the same parameters and under the same conditions as described for the macrojunction. With respect to fabrication of OLED nanopixels the following additional informations may be helpful. mCP was purchased from Ossila at unsublimed grade and was further purified by gradient sublimation. TXO-TPA was purchased from Lumtec at sublimed grade and used without further purification as TADF emitter in our devices. BPhen was purchased from Sigma Aldrich for spectrophotometric detection at a purity > 99.0 % and used without further purification. 5 nm HAT-CN was deposited at a deposition rate of 2 $nm \cdot min^{-1}$ and 30 nm NPB was deposited at a deposition rate of 4 $nm \cdot min^{-1}$. 30 nm of emissive layer were deposited by co-evaporation of mCP and TXO-TPA from separate crucibles. The ratio of deposition rates was adjusted to 7 vol % TXO-TPA doping with an overall deposition rate of 7 $nm \cdot min^{-1}$. 75 nm of Bphen were deposited at a deposition rate of 6 $nm \cdot min^{-1}$. After completion of organic deposition, the second shadow mask was aligned as described above. OLED top contact deposition was carried out using a glovebox-integrated deposition system from Leybold. 10 nm Ca were deposited at a rate of 0.2 to 0.3 $Å \, s^{-1}$ and 120 nm Al were deposited at a rate of 2 to 3 $Å \cdot s^{-1}$. During top contact deposition the sample was rotated at 16 rpm. Afterwards encapsulation was carried out inside the glovebox as described above, without further ambient exposure. For a better visualization, a simplified process sketch is shown in **Figure S8**.

*Atomic Force Microscopy:* Atomic force microscopy (AFM) height images were recorded in tapping mode at a Veeco Dimension Icon with standard tapping mode AFM probes (NanoWorld, NCHR, 320 kHz, 42 N·m$^{-1}$). Conductive AFM (C-AFM) images were recorded using the same machine in contact mode with the TUNA application module and conductive AFM probes (NanoWorld, SCM-PIC, 13 kHz, 0.2 N·m$^{-1}$). The measured electrodes were grounded, and a bias voltage of 100 mV was applied. Data analysis was carried out in Gwyddion.[43]

*(Opto-)electronic Device Characterization*: Current density-voltage characteristics of hole-only macrojunctions were recorded with a B1500A semiconductor parameter analyzer (Keysight Technologies). Contact was made through tungsten probe needles (FormFactor Inc.) which were mounted to triaxial probe arms (FormFactor Inc.) and micromanipulators (DPP210, FormFactor Inc.). In case of nanojunctions, voltages were applied by a source meter unit Keithley 2636B (Keithley Instruments Inc.). Contact was made through copper-beryllium probe needles (Semprex Corp.) which were mounted to triaxial probe arms (CascadeMicrotech) and micromanipulators (DPP 220, CascadeMicrotech). The current density-voltage characteristics of hole-only nanojunctions were measured in voltage steps of 100 mV with an integration time of 20 ms. The source limit was set to 100 nA. Electroluminescence (EL) of the nano-OLED device was collected by an oil-immersion microscope objective (Plan-Apochromat, 100x, NA = 1.45, Nikon) and detected through a spectrometer (Shamrock 303i, 80 lines·mm$^{-1}$, blazing at 600 nm or mirror) by an electron-multiplied charge-coupled device (iXon A-DU897-DC-BVF, Andor, EM gain 100). Source meter unit and EMCCD camera have been synchronized by a LabVIEW program to allow for a correlated data analysis. Consecutive constant voltage steps were applied for measuring the voltage-dependent spectra. Typical integration times for recording the spectra ranged from 0.1 s to 1 s. The external quantum efficiency was calculated from the EL spectra considering the detection efficiency of the optical setup.

## Supporting Information

Supporting Information is available from the Wiley Online Library or from the author.

## Acknowledgements

C. Zhang and B. Ewald contributed equally to this manuscript as joined first authors. The German Research Foundation (projects HE 5618/8-1 and PF385/12-1) is acknowledged for

WILEY-VCH

financial support. J.P. appreciates financial support by the Bavarian State Ministry of Science, Research, and the Arts (Collaborative Research Network "Solar Technologies Go Hybrid").

**Conflict of Interest**

The authors declare no conflict of interest.

**Data Availability Statement**

The data that support the findings of this study are available from the corresponding authors upon reasonable request.

**References**

[1]     W. Brütting, S. Berleb, A.G. Mückl, *Org. Electron.* **2001**, *2*, 1, DOI: https://doi.org/10.1016/S1566-1199(01)00009-X.

[2]     A. M. T. Muthig, O. Mrozek, T. Ferschke, M. Rödel, B. Ewald, J. Kuhnt, C. Lenczyk, J. Pflaum, A. Steffen, *J. Am. Chem. Soc.* **2023**, *145*, 4438, DOI: https://doi.org/10.1021/jacs.2c09458.

[3]     S. Reineke, M. Thomschke, B. Lüssem, K. Leo, *Rev. Mod. Phys.* **2013**, *85*, 1245, DOI: https://doi.org/10.1103/RevModPhys.85.1245.

[4]     D. Yang, D. Ma, *Adv. Opt. Mater.* **2019**, *7*, 1800522, DOI: https://doi.org/10.1002/adom.201800522.

[5]     H. Kleemann, K. Krechan, A. Fischer, K. Leo, *Adv. Funct. Mater.* **2020**, *30*, 1907113, DOI: https://doi.org/10.1002/adfm.201907113.

[6]     H. S. Wasisto, J. D. Prades, J. Gülink, A. Waag, *Appl. Phys. Rev.* **2019**, *6*, 041315, DOI: https://doi.org/10.1063/1.5096322.

[7]     W.-J. Joo, J. Kyoung, M. Esfandyarpour, S.-H. Lee, H. Koo, S. Song, Y.-N. Kwon, S. H. Song, J. C. Bae, A. Jo, *Science* **2020**, *370*, 459, DOI: https://doi.org/10.1126/science.abc8530.

[8]     F. Mariani, I. Gualandi, W. Schuhmann, E. Scavetta, *Microchim. Acta* **2022**, *189*, 459, DOI: https://doi.org/10.1007/s00604-022-05548-3.

[9]     U. Zschieschang, U. Waizmann, J. Weis, J. W. Borchert, H. Klauk, *Sci. Adv.* **2022**, *8*, eabm9845, DOI: https://doi.org/10.1126/sciadv.abm9845.

[10]    C. Eckel, J. Lenz, A. Melianas, A. Salleo, R. T. Weitz, *Nano Lett.* **2022**, *22*, 973, DOI: https://doi.org/10.1021/acs.nanolett.1c03832.

[11]    A. Ghosh, B. Corves, *Introduction to Micromechanisms and Microactuators*, Springer, **2015**.

[12]    F. A. Boroumand, P. W. Fry, D. G. Lidzey, *Nano Lett.* **2005**, *5*, 67, DOI: https://doi.org/10.1021/nl048382k.

[13]    H. Yamamoto, J. Wilkinson, J. P. Long, K. Bussman, J. A. Christodoulides, Z. H. Kafafi, *Nano Lett.* **2005**, *5*, 2485, DOI: https://doi.org/10.1021/nl051811+.

[14]    J. G. Wilbers, B. Xu, P. A. Bobbert, M. P. de Jong, W.G. van der Weil, *Sci. Rep.* **2017**, *7*, 41171, DOI: https://doi.org/10.1038/srep41171.

[15]    G. Smit, S. Rogge, T. Klapwijk, *Appl. Phys. Lett.* **2002**, *80*, 2568, DOI: https://doi.org/10.1063/1.1467980.

[16]    G. Smit, S. Rogge, T. Klapwijk, *Appl. Phys. Lett.* **2002**, *81*, 3852, DOI: https://doi.org/10.1063/1.1521251.

[17]    M. Rezeq, K. Eledlebi, M. Ismail, R. K. Dey, B. Cui, *J. Appl. Phys.* **2016**, *120*, 044302, DOI: https://doi.org/10.1063/1.4959090.

[18] Y. Shen, N. C. Giebink, *Phys. Rev. Appl.* **2015**, *4*, 054017, DOI: https://doi.org/10.1103/PhysRevApplied.4.054017.

[19]    W.-J Joo, T.-L. Choi, J. Lee, S. K. Lee, M.-S. Jung, N. Kim, J. M. Kim, *J. Phys. Chem. B* **2006**, *110*, 23812, DOI: https://doi.org/10.1021/jp0649899.

[20]    S. Gao, C. Song, C. Chen, F. Zeng, F. Pan, *J. Phys. Chem. C* **2012**, *116*, 17955, DOI: https://doi.org/10.1021/jp305482c.

[21]    Z. Wang, F. Zeng, J. Yang, C. Chen, F. Pan, *ACS Appl. Mater. Interfaces.* **2012**, *4*, 447, DOI: https://doi.org/10.1021/am201518v.

[22]    L. Novotny, B. Hecht, *Principles of Nano-Optics*, Cambridge University Press, **2012**.

[23]    M. A. Fussella, R. Saramak, R. Bushati, V. M. Menon, M. S. Weaver, N. J. Thompson, J. J. Brown, *Nature* **2020**, *585*, 379, DOI: https://doi.org/10.1038/s41586-020-2684-z.

[24]    B. Munkhbat, H. Pöhl, P. Denk, T. A. Klar, M. C. Scharber, C. Hrelescu, *Adv. Opt. Mater.* **2016**, *4*, 772, DOI: https://doi.org/10.1002/adom.201500702.

[25]    Y. Qu, M. Slootsky, S. R. Forrest, *Nat. Photonics* **2015**, *9*, 758, DOI: https://doi.org/10.1038/nphoton.2015.194.

[26]     J. Frischeisen, Q. Niu, A. Abdellah, J. B. Kinzel, R. Gehlhaar, G. Scarpa, C. Adachi, P. Lugli, W. Brütting, *Opt. Express* **2011**, *19*, A7, DOI: https://doi.org/10.1364/OE.19.0000A7.

[27]     P. Grimm, S. Zeißner, M. Rödel, S. Wiegand, S. Hammer, M. Emmerling, E. Schatz, R. Kullock, J. Pflaum, B. Hecht, *Nano Lett.* **2022**, *22*, 1032, DOI: https://doi.org/10.1021/acs.nanolett.1c03994.

[28]     M. Ochs, L. Jucker, M. Rödel, M. Emmerling, R. Kullock, J. Pflaum, M. Mayor, B. Hecht, *Nanoscale* **2023**, *15*, 5249, DOI: https://doi.org/10.1039/D2NR06343C.

[29]     P. Pertsch, R. Kullock, V. Gabriel, L. Zurak, M. Emmerling, B. Hecht, *Nano Lett.* **2022**, *22*, 6982, DOI: https://doi.org/10.1021/acs.nanolett.2c01772.

[30]     R. Kullock, M. Ochs, P. Grimm, M. Emmerling, B. Hecht, *Nat. Commun.* **2020**, *11*, 115, DOI: https://doi.org/10.1038/s41467-019-14011-6.

[31]     M. Ochs, L. Zurak, E. Krauss, J. Meier, M. Emmerling, R. Kullock, B. Hecht, *Nano Lett.* **2021**, *21*, 4225, DOI: https://doi.org/10.1021/acs.nanolett.1c00182.

[32]     M. J. Word, I. Adesida, P. R. Berger, *J. Vac. Sci. Technol.* **2003**, *21*, L12, DOI: https://doi.org/10.1116/1.1629711.

[33]     T. Ferschke, A. Hofmann, W. Brütting, J. Pflaum, *ACS Appl. Electron. Mater.* **2019**, *2*, 186, DOI: https://doi.org/10.1021/acsaelm.9b00687.

[34]     A. Kahn, *Mater. Horiz.* **2016**, *3*, 7, DOI: https://doi.org/10.1039/C5MH00160A.

[35]     E. Oh., S. Park, J. Jeong, S. J. Kang, H. Lee, Y. Yi, *Chem. Phys. Lett.* **2017**, *668*, 64, DOI: https://doi.org/10.1016/j.cplett.2016.12.007.

[36]     Y.-K. Kim, J. Won Kim, Y. Park, *Appl. Phys. Lett.* **2009**, *94*, 063305, DOI: https://doi.org/10.1063/1.3081409.

[37]     A. Kahn, N. Koch, W. Gao, *J. Polym. Sci. B: Polym. Phys.* **2003**, *41*, 2529, DOI: https://doi.org/10.1002/polb.10642.

[38]     N. Turetta, F. Sedona, A. Liscio, M. Sambi, P. Samorì, *Adv. Mater. Interfaces* **2021**, *8*, 2100068, DOI: https://doi.org/10.1002/admi.202100068.

[39]     J. Van Der Holst, M. Uijttewaal, B. Ramachandhran, R. Coehoorn, P. Bobbert, G. De Wijs, R. De Groot, *Phys. Rev. B* **2009**, *79*, 085203, DOI: https://doi.org/10.1103/PhysRevB.79.085203.

[40]     A. Köhler, H. Bässler, *Electronic Processes in Organic Semiconductors: An Introduction*, John Wiley & Sons, **2015**.

[41]     A. Fleissner, H. Schmid, C. Melzer, H. von Seggern, *Appl. Phys. Lett.* **2007**, *91*, 242103, DOI: https://doi.org/10.1063/1.2820448.

[42]    H. Wang, L. Xie, Q. Peng, L. Meng, Y. Wang, Y. Yi, P. Wang, *Adv. Mater.* **2014**, *26*, 5198, DOI: https://doi.org/10.1002/adma.201401393.

[43]    D. Nečas, P. Klapetek, *Open Phys.* **2012**, *10*, 181, DOI: https://doi.org/10.2478/s11534-011-0096-2.

**Individually Addressable Nanoscale OLEDs**

*Cheng Zhang\*, Björn Ewald\*, Leo Siebigs, Luca Steinbrecher, Maximilian Rödel, Monika Emmerling, Jens Pflaum\*, Bert Hecht\**

This work introduces a novel concept for creating individually addressable nano-OLEDs using nanoscale electrodes that support plasmonic modes. By selectively insulating sharp electrode contours, we achieve controlled charge-carrier injection and suppress failure mechanisms. Demonstrating stable and efficient operation of $300 \times 300 \ nm^2$ OLED pixels, this approach enables further miniaturization and offers potential for future improvements using plasmonic effects.

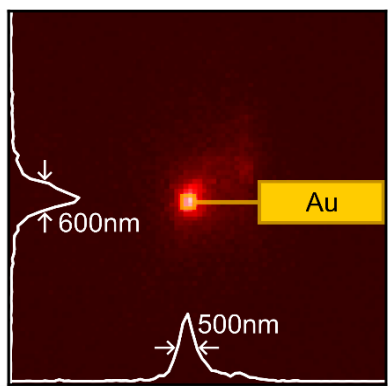

Supporting Information

**Individually Addressable Nanoscale OLEDs**

*Cheng Zhang\*, Björn Ewald\*, Leo Siebigs, Luca Steinbrecher, Maximilian Rödel, Monika Emmerling, Jens Pflaum\*, Bert Hecht\**

**SI 1: Electrostatic Simulations**

When a macroscopic extended electrode (in our case anode) is scaled down to a nanoscale electrode with subwavelength dimensions, electric field inhomogeneities will dominate the overall device operation in a vertical device architecture. As sketched in Figure S1a, the static electric field is expected to be locally intensified at the edges of the electrodes when a DC voltage bias is applied, whereas uniform field lines are expected to be found at the extended top planar electrode. An electrostatic simulation was performed using COMSOL Multiphysics 6.0, AC/DC module, to demonstrate this effect, as depicted in Figure S1b and c. The electric field distribution beneath the extended planar top cathode is indeed uniform at 1 nm distance, but there is a notably enhanced inhomogeneous electric field at the corners (> 6 times enhancement) and edges (> 3 times enhancement) of the structured bottom electrode, compared to the planar top electrode. Such electric field inhomogeneities lead to localized edge-induced charge injection and enhanced electromigration causing filament growth.

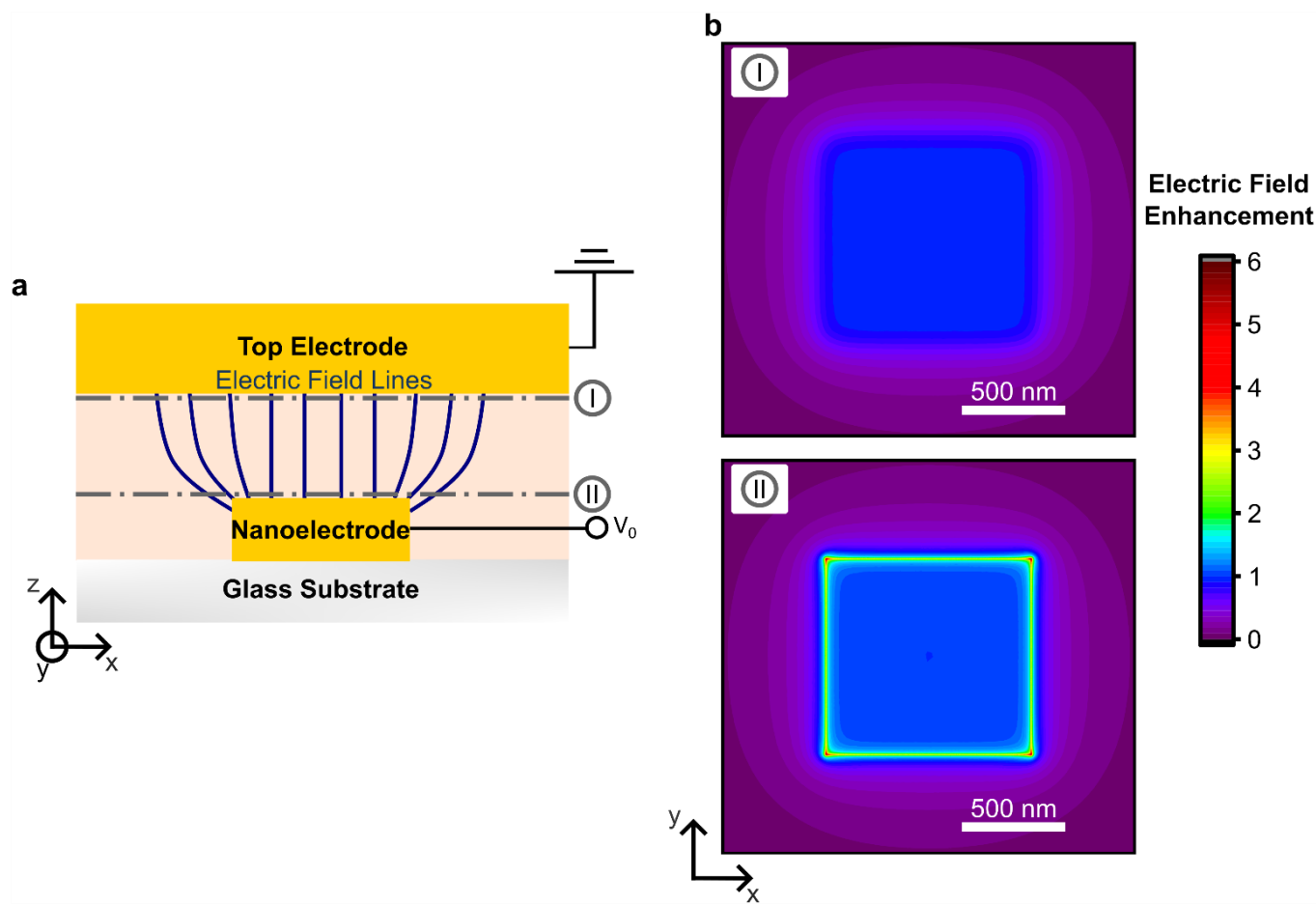

**Figure S1.** Influence of electric field inhomogeneity in a vertical organic optoelectronic device using a structured bottom electrode. a) Sketch of the electrostatic field distribution in a vertical metal-organic-metal device. The structured Au anode is voltage biased, while the flat Au electrode is grounded. b) Electrostatic field distribution 1 nm below the flat extended Au cathode and 1 nm above the $1 \times 1$ μm$^2$ structured Au anode (both plains marked by dashed dotted lines in a)). To obtain a realistic estimate of the local field enhancement all corners and edges are rounded with a 10 nm radius of curvature. The distance between two electrodes is 140 nm. The organic layer between the electrodes is modelled by a dielectric (relative permittivity $\varepsilon_r$ = 2.25). A voltage of 10 V is applied between the electrodes. The absolute electric field is normalized to the bottom electrode center.

## SI 2: Morphological Characterization of Au Thin Films

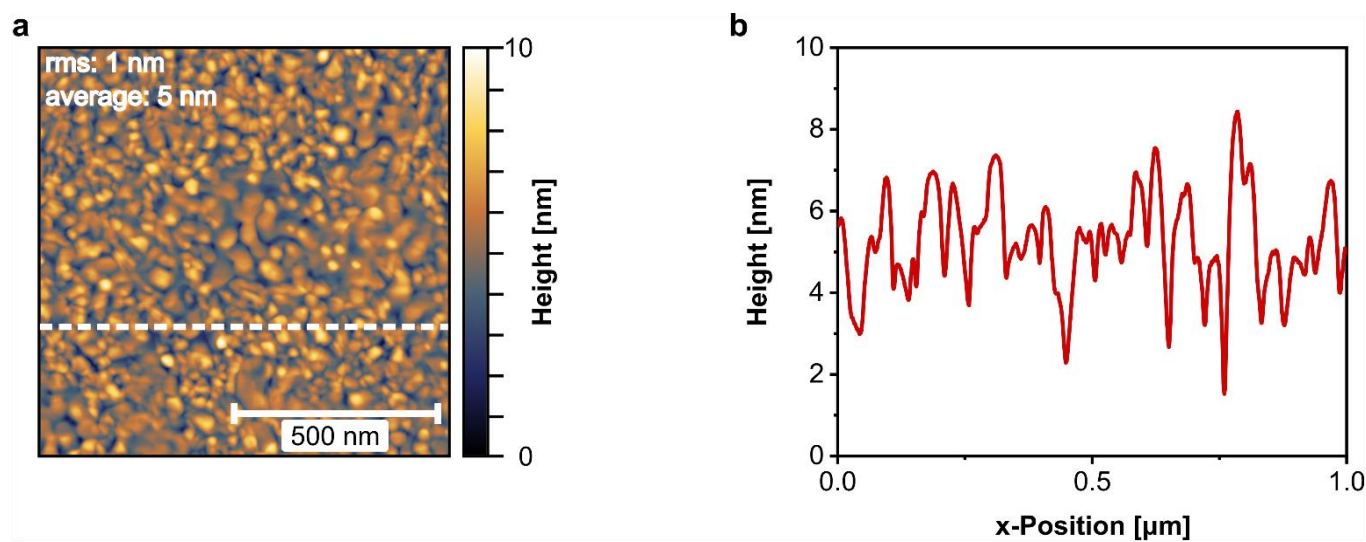

**Figure S2.** Morphology of Au thin films (thickness 50 nm) deposited at a moderate rate of 1.5 nm·s⁻¹: a) Tapping mode AFM image of a representative Au film area (1 × 1 µm²) with a root mean square (rms) roughness of 1 nm and an average roughness of 5 nm. b) Height cross section along the white dashed line in a.

## SI 3: HAT-CN Functionalization of Au Anodes

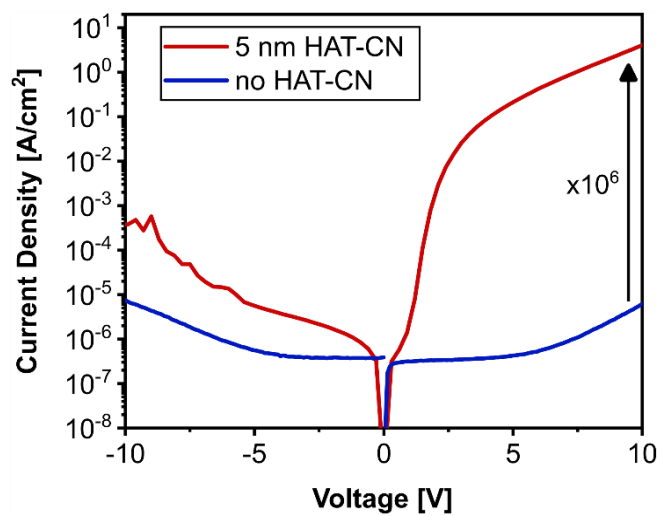

**Figure S3.** Current density-voltage characteristics of hole-only macrojunctions (100 × 100 µm²) with 5 nm HAT-CN functionalization and without HAT-CN functionalization. The hole current density at 10 V is increased by 6 orders of magnitude upon HAT-CN functionalization. The device architectures are 50 nm Au / 5 nm HAT-CN / 135 nm NPB / 140 nm Au and 50 nm Au / 140 nm NPB / 140 nm Au, respectively.

**SI 4: Absolute Current in Hole-Only Macro- and Nanojunctions**

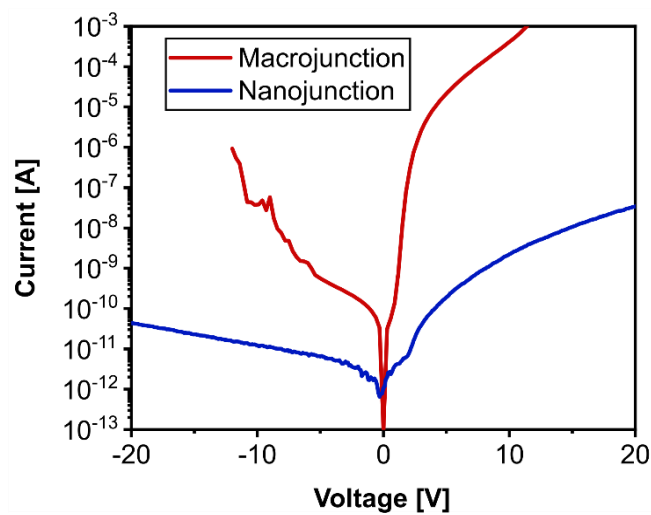

**Figure S4.** Current-voltage characteristics of a hole-only macrojunction (electrode patch: $100 \times 100$ µm², active area: $1.0 \cdot 10^{-4}$ cm²) and nanojunction (electrode patch: $1 \times 1$ µm², nanoaperture diameter: 550 nm, active area: $2.4 \cdot 10^{-9}$ cm²) in semilogarithmic presentation. The absolute current at 10 V is 5 orders of magnitude smaller in the nanojunction (10 nA) compared to the macrojunction (1 mA).

**SI 5: Influence of Nanoaperture Diameter on the Blocking Ratio**

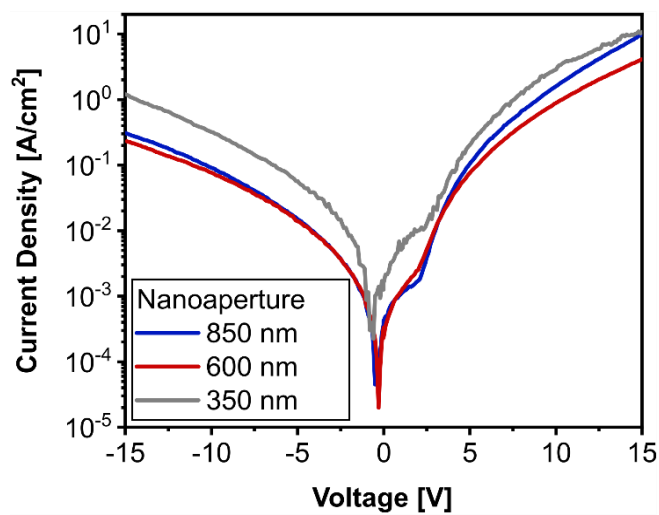

**Figure S5.** Current density-voltage characteristics of three hole-only nanojunctions (electrode patch: $1 \times 1$ µm$^2$) with varying nanoaperture diameter (850 nm. 600 nm, 350 nm) and a fixed depth of 50nm in semilogarithmic presentation. The blocking ratio @15 V (forward to reverse current density) decreases with decreasing nanoaperture diameter from 33 (850 nm) to 17 (600 nm) and finally to 8 (350 nm). We attribute this effect to a slight increase in the top contact curvature imposed by the nanoaperture shape.

**SI 6: Hole-Only Nanojunctions with 300 × 300 nm² Au nanoelectrodes**

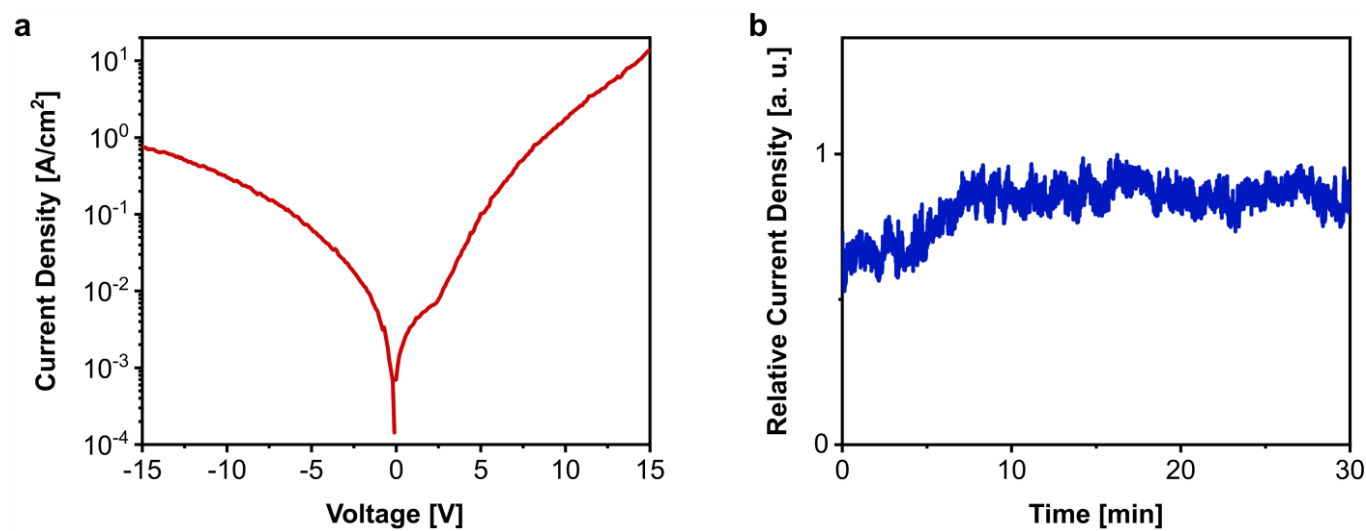

**Figure S6.** Hole-only nanojunctions with an electrode size of 300 × 300 nm² and a nanoaperture opening of 200 nm in diameter: a) Current density-voltage characteristics of a representative nanojunction pixel in semilogarithmic presentation. b) Constant voltage operation (10 V, dc) of the nanojunction. The current density is stable over the observation period of 30 minutes and no tendency for filament formation is observed.

**SI 7: Stability and External Quantum Efficiency of OLED Nanopixels**

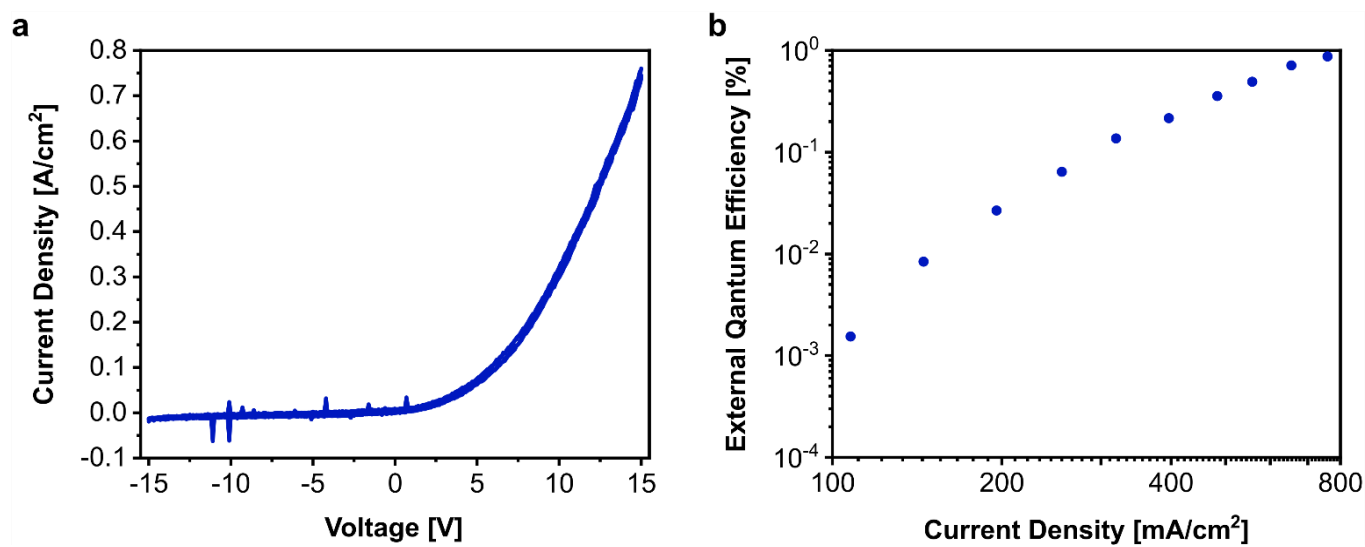

**Figure S7.** Stability and external quantum efficiency of OLED nanopixels: a) Current density-voltage characteristics of an OLED nanopixel. The pixel has been cycled three times forth and back starting from 0 V. b) External quantum efficiency as function of the current density.

**SI 8: Fabrication Workflow of Hole-Only Nanojunctions**

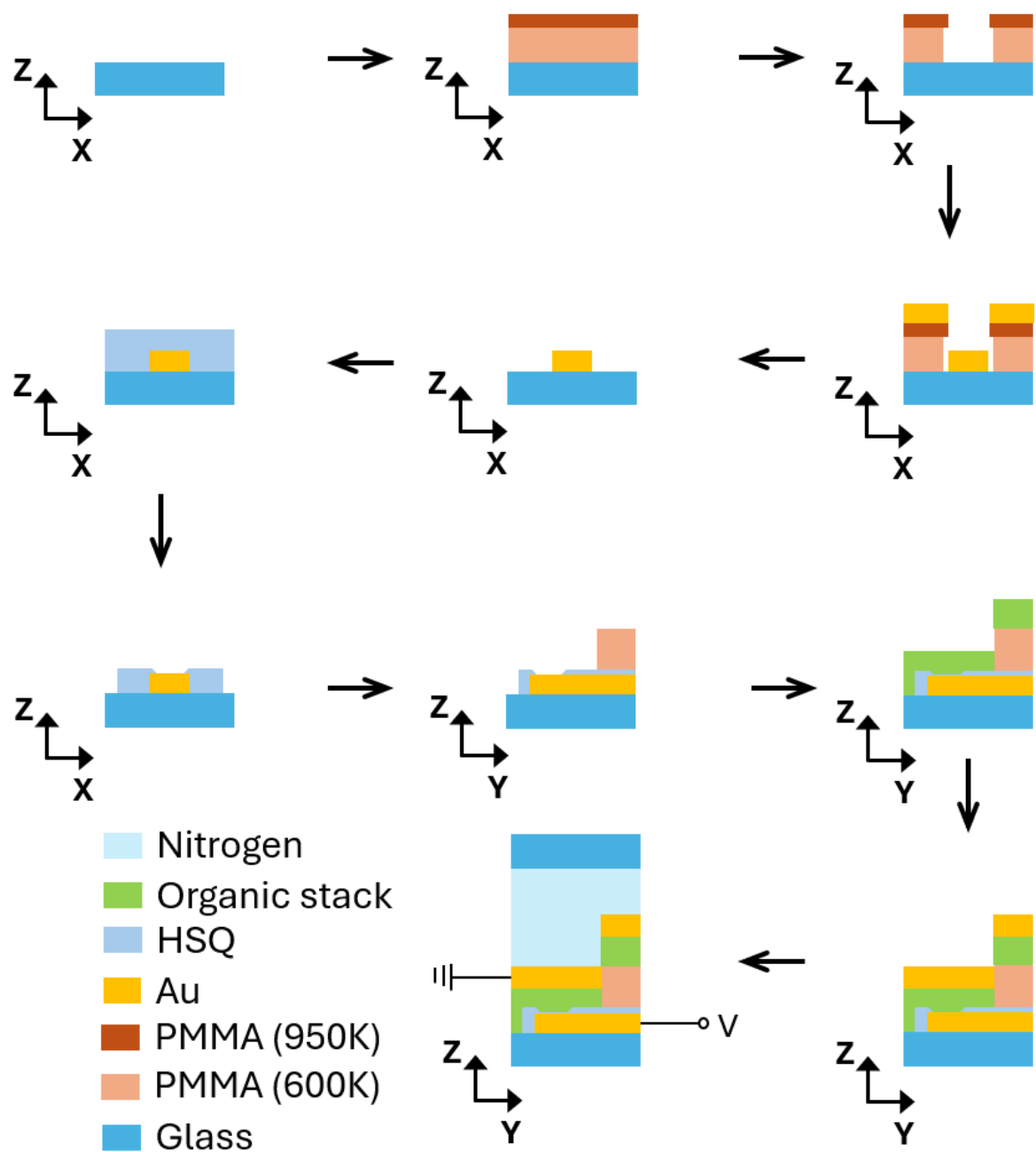

**Figure S8.** Fabrication workflow for a hole-only nanojunction. 1. Electrode layout fabrication; 2. Double PMMA layer spincoating; 3. 1$^{st}$ electron-beam writing to pattern the antenna; 4. Thermal evaporation of Au; 5. Lift-off process; 6. HSQ spincoating; 7. Nanoaperture fabrication via the 2$^{nd}$ electron-beam writing; 8. 3$^{rd}$ electron-beam writing to define the pixel with a PMMA insulating layer; 9. Organic stack deposition via stencil lithography; 10. Top metal contact deposition via stencil lithography; 11. Device encapsulation with epoxy resin in a glovebox system.